\documentclass[12pt]{iopart}
\usepackage{graphicx}
\usepackage{epsfig}

\begin{document}

\title{Percolation of localized attack on complex networks}

\author{Shuai Shao\dag
\footnote[3]{To
whom correspondence should be addressed (sshao@bu.edu)},
 Xuqing Huang\dag, H. Eugene Stanley\dag, and Shlomo Havlin\dag\ddag}
 
 \address{\dag\ Center for Polymer Studies and Department of Physics,
Boston University, Boston, MA 02215, USA}
\address{\ddag\ Department of Physics,
Bar-Ilan University, Ramat-Gan 52900, Israel}

\begin{abstract}
{The robustness of complex networks against node failure and malicious attack has been of interest for decades, while most of the research has focused on random attack or hub-targeted attack. In many real-world scenarios, however, attacks are neither random nor hub-targeted, but localized, where a group of neighboring nodes in a network are attacked and fail. In this paper we develop a percolation framework to analytically and numerically study the robustness of complex networks against such localized attack. In particular, we investigate this robustness in Erd\H{o}s-R\'{e}nyi networks, random-regular networks, and scale-free networks.  Our results provide insight into how to better protect networks, enhance cybersecurity, and facilitate the design of more robust infrastructures.}
\end{abstract}

\maketitle



The functioning of complex networks such as the Internet, airline
routes, and social networks is crucially dependent upon the
interconnections between network nodes. These interconnections are such
that when some nodes in the network fail, others connected through them
to the network will also be disabled and the entire network may
collapse. In order to understand network robustness and design resilient
complex systems, one needs to know whether a complex network can continue
to function after a fraction of its nodes have been removed either
through node failure or malicious attack \cite{Albert2000, Cohen2000,
  Callaway2000, Cohen2001, Barabasi2002,
  Imre2005, Gallos2005, Newmanbook, Bashan2011, Sergey2010, Parshani2010, Huang2011, 
  Bashan2012, Jianxi2011, Jianxi2012, Jianxi2013, Brummitt2012, Baxter2012, Peixoto2012, Cohenbook}. This question is dealt with in percolation
theory~\cite{Cohenbook, Bundebook, Staufferbook, Coniglio} in which the
percolation phase transition occurs at some critical occupation
probability $p_c$. Above $p_c$, a giant component, defined as a cluster
whose size is proportional to that of the entire network, exists; below
$p_c$ the giant component is absent and the entire network
collapses. Only nodes in the giant component continue to function after
the node-removal process.

The robustness of complex networks under attack is dependent upon the
structure of the underlying network and the nature of the
attack. Previous research has focused on two types of initial attack:
random attack and hub-targeted attack. In a random attack each node in
the network is attacked with the same probability~\cite{Albert2000,
  Cohen2000, Callaway2000, Newmanbook, Cohenbook, Sergey2010}. In a hub-targeted
attack the probability that high-degree nodes will be attacked is higher
than that for low-degree nodes~\cite{Albert2000, Callaway2000,
  Cohen2001, Gallos2005, Huang2011}. An important feature of the network
structure is its degree distribution, $P(k)$, which describes the
probability that each node has a specific degree $k$. Networks with
different degree distributions behave very differently under different
types of attack. For instance, the Internet, which shows a power law
degree distribution, is extremely robust against random attack but
vulnerable to hub-targeted attack~\cite{Albert2000, Cohen2000}.

\begin{figure}
\begin{center}
\includegraphics[scale = 0.5]{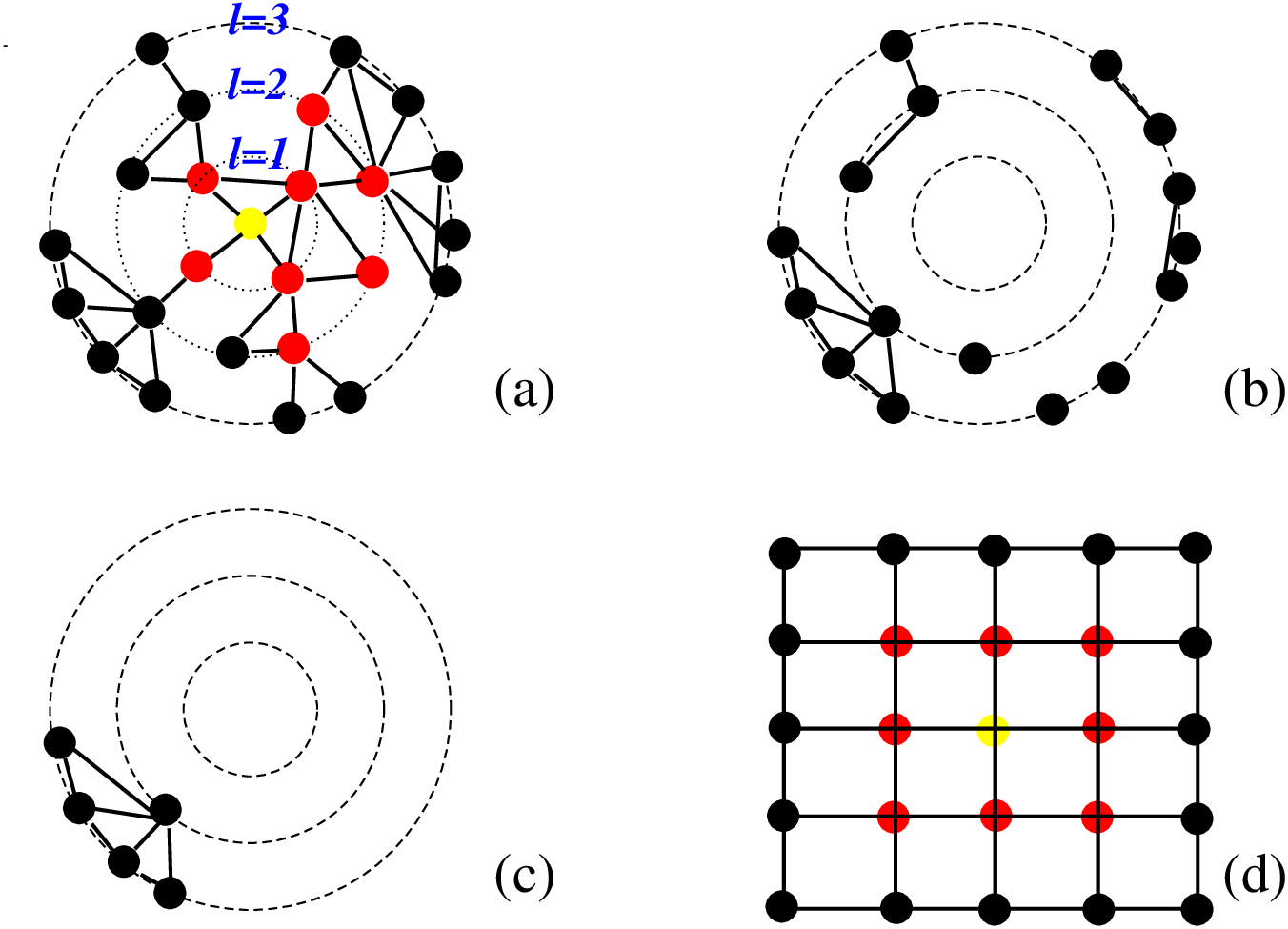}
\caption{Schematic illustration of the localized attack process. (a) A fraction $1-p$ of the nodes are chosen to be removed, starting from the root node, its nearest neighbors, next nearest neighbors, and so on (yellow represents the root node, red the other nodes to be removed). (b) Remove the chosen nodes and the links. An attacked ``hole'' centered around the root node is formed. (c) Only nodes in the giant component (largest cluster) keep functioning and are left in the network.  (d) Localized attack on regular lattice (here, square lattice). For a regular lattice with $N\rightarrow\infty$, one needs to attack all nodes in order to collapse the network, i.e., $p_c\rightarrow 0$.} 
\label{demo}
\end{center}
\end{figure}

However these two types of attack---random attack and hub-targeted
attack---do not adequately describe many real-world scenarios in which
complex networks suffer from damage that is localized, i.e., a node is
affected, then its neighbors, and then their neighbors, and so on (see
Fig.~\ref{demo}). Examples include the effects of earthquakes, floods,
or military attacks on infrastructure networks and the effects of a
computer virus or malware on computer networks. Recent occurrences of
the latter include attacks carried out by cybercriminals who create a
``botnet'', a cluster of neighboring ``zombie computers'' in a computer
network and, by using them, are able to damage the entire network. An
understanding of the effect of this kind of attack on the functioning of
a network is still lacking.

Here we will analyze the robustness of complex networks sustaining this
kind of localized attack in order to determine how much damage a network
can sustain before it collapses, i.e., to find the percolation threshold
$p_c$. We also want to predict the fraction of nodes that keep
functioning after an initial attack of a fraction of $1-p$ nodes, i.e.,
the relative size of the giant component (the order parameter),
$P_{\infty}$. Note that localized attack has been studied only on
specific network structures \cite{Neumayer2009} or on interdependent
spatially embedded networks \cite{Berezin2013}, but a general
theoretical formalism for studying localized attacks on complex networks
is currently missing.

Here we develop a mathematical framework for studying localized attacks
on complex networks with arbitrary degree distribution and we find exact
solutions for percolation properties such as the critical threshold
$p_c$ and the relative size of the giant component $P_{\infty}$. In
particular, we apply our framework to study and compare the robustness
of three types of random networks, (i) Erd\H{o}s-R\'{e}nyi (ER) networks
with a Poissonian degree distribution ($P(k)=e^{-\langle
  k\rangle}\langle k\rangle^k/k!$)~\cite{BollobasBook}, (ii)
random-regular (RR) networks with a Kronecker delta degree distribution
($P(k)=\delta_{k,k_0}$), and (iii) scale-free (SF) networks with a power
law degree distribution ($P(k)\sim
k^{-\lambda}$)~\cite{Barabasi2002}. We find that the effect of a
localized attack on an ER network is identical to that of a random
attack. For an RR network, we find that the $p_c$ of a localized attack
is always smaller (i.e., more robust) than that of a random
attack. However, the robustness of a SF network against localized attack
is found to be critically dependent upon the power law exponent
$\lambda$. Surprisingly, a critical exponent $\lambda_c$ exists such
that when $\lambda<\lambda_c$, for localized attack the network is
significantly more vulnerable compared to random attack, with $p_c$ being
larger. While for $\lambda>\lambda_c$, the opposite is true.

Consider a random network with a degree distribution $P(k)$, which
indicates the probability that a node in the network has $k$
neighbors. The generating function of the degree distribution is defined
as $G_0(x)=\sum_{k=0}^{\infty}{P(k)x^k}$~\cite{Newman2001, Molly1995}.
We start from a randomly chosen ``root'' node. All nodes in the random
network are listed in ascending order of their distances from this root
node (see Fig.~\ref{demo}(a)). The shell $l$ is defined as the set of
nodes that are at distance $l$ from the root node~\cite{Kalisky2006,
  JiaShao2009}. Within the same shell, all nodes are at the same
distance from the root node and are positioned randomly.

We initiate the localized attack process by removing the root node, then
the nodes in the first shell, and so on. We remove nodes in the
ascending order of their distances from the root node. Within the same
shell we remove nodes randomly and, after nodes in shell $l$ are fully
removed, we begin removing nodes in shell $l+1$. We continue the
localized attack process until a fraction $1-p$ of nodes in the entire
network are removed. Thus a ``hole'' of attacked nodes forms around the
root node. The remaining $p$ fraction of nodes in the network are those
at greater distances from the root node (see Fig.~\ref{demo}(b)). After
the initial removal of $1-p$ fraction of the network nodes and all links
connected to them, the remaining network fragments into connected
clusters. As in percolation theory~\cite{Bundebook, Staufferbook}, only
nodes in the giant component (the largest cluster) are still
functional. Nodes belonging to other small clusters are considered
non-functional and are also removed (see Fig.~\ref{demo}(c)). Note that
for localized attack on a regular lattice, as the number of network
nodes $N\rightarrow\infty$, $p_c\rightarrow 0$, i.e., one has to attack
all nodes in the regular lattice in order to collapse the lattice (see
Fig.~\ref{demo}(d)).

We find that the generating function of the degree distribution of the
remaining network after the localized attack is (see supplementary
information)
\begin{equation}
G^{p}_{0}(x)=\frac{1}{G_0(f)}G_0[f+\frac{G^{'}_0(f)}{G^{'}_0(1)}(x-1)], 
\label{gpx}
\end{equation}
where $p$ is the fraction of unremoved nodes and $f\equiv
G^{-1}_0(p)$. The critical probability $p_c$ where the network collapses
and the size of the giant component $P_{\infty}(p)$ for $p>p_c$ can be
derived analytically from Eq.~(\ref{gpx}). The generating function of
the cluster sizes in the remaining network is
$H^{p}_{0}(x)=xG^{p}_{0}(H^{p}_{1}(x))$, where $H^{p}_{1}(x)$ satisfies
the transcendental equation $H^{p}_{1}(x)=xG^{p}_{1}(H^{p}_{1}(x))$ and
$G^{p}_{1}(x)=G^{'p}_{0}(x)/G^{'p}_{0}(1)$~\cite{Newman2001}. By
combining Eq.~(\ref{gpx}) and the criterion for the network to
collapse~\cite{Cohen2000, Callaway2000}, $G^{'p}_{1}(1)=1$, we find
that $p_c$ satisfies
\begin{equation}
G_0^{''}(G_0^{-1}(p_c))=G_0^{'}(1).\label{fclocal} 
\end{equation}
The size of the giant component $S(p)$ as a fraction of the remaining
network satisfies 
\begin{equation}
S(p)=1-G^{p}_{0}(H^{p}_{1}(1)), \label{gcremain} 
\end{equation}
where $H^{p}_{1}(1)$ satisfies
$H^{p}_{1}(1)=G^{p}_{1}(H^{p}_{1}(1))$. The relative size of the giant
component as a fraction of the original network is
$P_{\infty}(p)=pS(p)$.

We apply the above mathematical framework to three types of complex
networks: Erd\H{o}s-R\'{e}nyi (ER) networks, random-regular (RR)
networks, and scale-free (SF) networks, and compare the results of a
localized attack with those of a random attack.

\begin{figure}
\begin{center}
\includegraphics[scale=0.5]{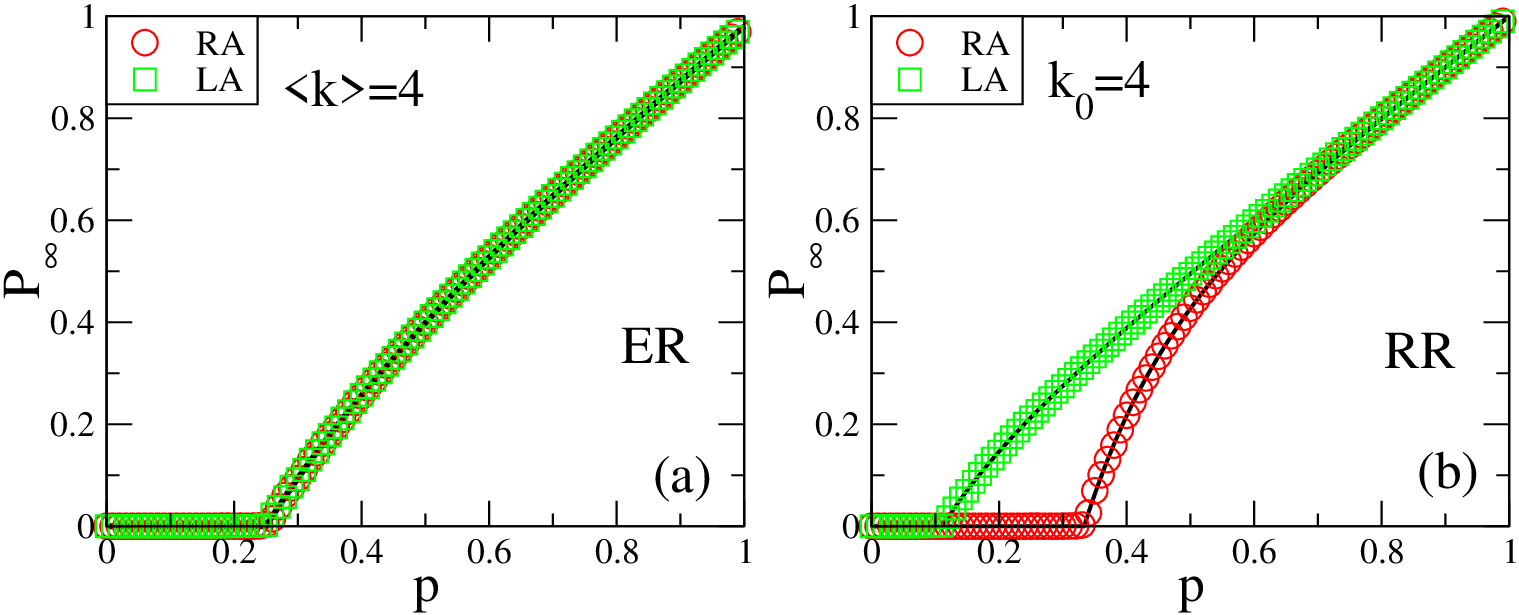}
\caption{Percolation transitions for (a) an ER network and (b) an RR network under localized attack (LA) and random attack (RA), with network size $N=10^6$, average degree $\langle k\rangle=4$ in ER network, and $k_0=4$ in RR network. Theoretical results (solid lines) and simulations (symbols)  agree well with each other. Note that the effect of localized attack and random attack on an ER network (see (a)) are identical (here, $p_c=1/\langle k\rangle=0.25$), while an RR network (see (b)) is more robust against localized attack compared to random attack.} \label{fig_gc_rrer}
\end{center}
\end{figure}

For an ER network with an average degree $\langle k\rangle$, the degree
distribution follows a Poissonian distribution $P(k)=e^{-\langle
  k\rangle}\langle k\rangle^k/k!$ and the corresponding generating
function of degree distribution is $G_0(x)=e^{\langle
  k\rangle(x-1)}$. From Eq.~(\ref{gpx}) we have
$G^{p}_{0}(x)=e^{p\langle k\rangle(x-1)}$, which is the same as the
generating function of the degree distribution for the remaining network
after a random attack. Thus the effect of a localized attack is exactly
the same as that of a random attack on an ER network (see
Fig.~\ref{fig_gc_rrer}(a)), and the critical threshold is $p_c=1/\langle
k\rangle$. The size of the giant component $P_{\infty}(p)$ satisfies
$P_{\infty}(p)=p(1-e^{-\langle k\rangle P_{\infty}(p)})$. In an RR
network each node is connected to $k_0$ other nodes randomly and the
generating function of the degree distribution is $G_0(x)=x^{k_0}$.
Using Eq.~(\ref{fclocal}) we find that the critical threshold for a
localized attack on an RR network is

\begin{equation}
p_c=(k_0-1)^{-\frac{k_0}{k_0-2}}. 
\end{equation}

Note that for an RR network under random attack the critical threshold
is $p_c=(k_0-1)^{-1}$. Thus, for $k_0>2$, $p_c$ under localized attack
is always smaller than $p_c$ under random attack (see
Fig.~\ref{fig_gc_rrer}(b)). This means that an RR network is more
resilient against localized attack than against random attack. When
$k_0\gg 1$, random and localized attacks have the same critical
threshold ($p_c=1/(k_0-1)$), since in this limit every node is a
neighbor of the root node and there is no difference between random and
localized attacks. Since $\lim_{k_0 \rightarrow 2}p_c=e^{-2}\approx
0.135$ and $\lim_{k_0 \rightarrow \infty}p_c=0$, one can see that $p_c$
for a localized attack on an RR network is always within the range
$(0,e^{-2})$ for all $k_0>2$. For $p>p_c$, from Eq.~(\ref{gcremain}),
the relative size of the giant component $P_\infty(p)$ satisfies
\begin{equation}
(p-P_\infty(p))^{\frac{1}{k_0}}-p^{\frac{1}{k_0}}=
(p-P_\infty(p))^{\frac{k_0-1}{k_0}}-p^{\frac{k_0-1}{k_0}}. 
\end{equation}

\begin{figure}
\begin{center}
\includegraphics[scale=0.5]{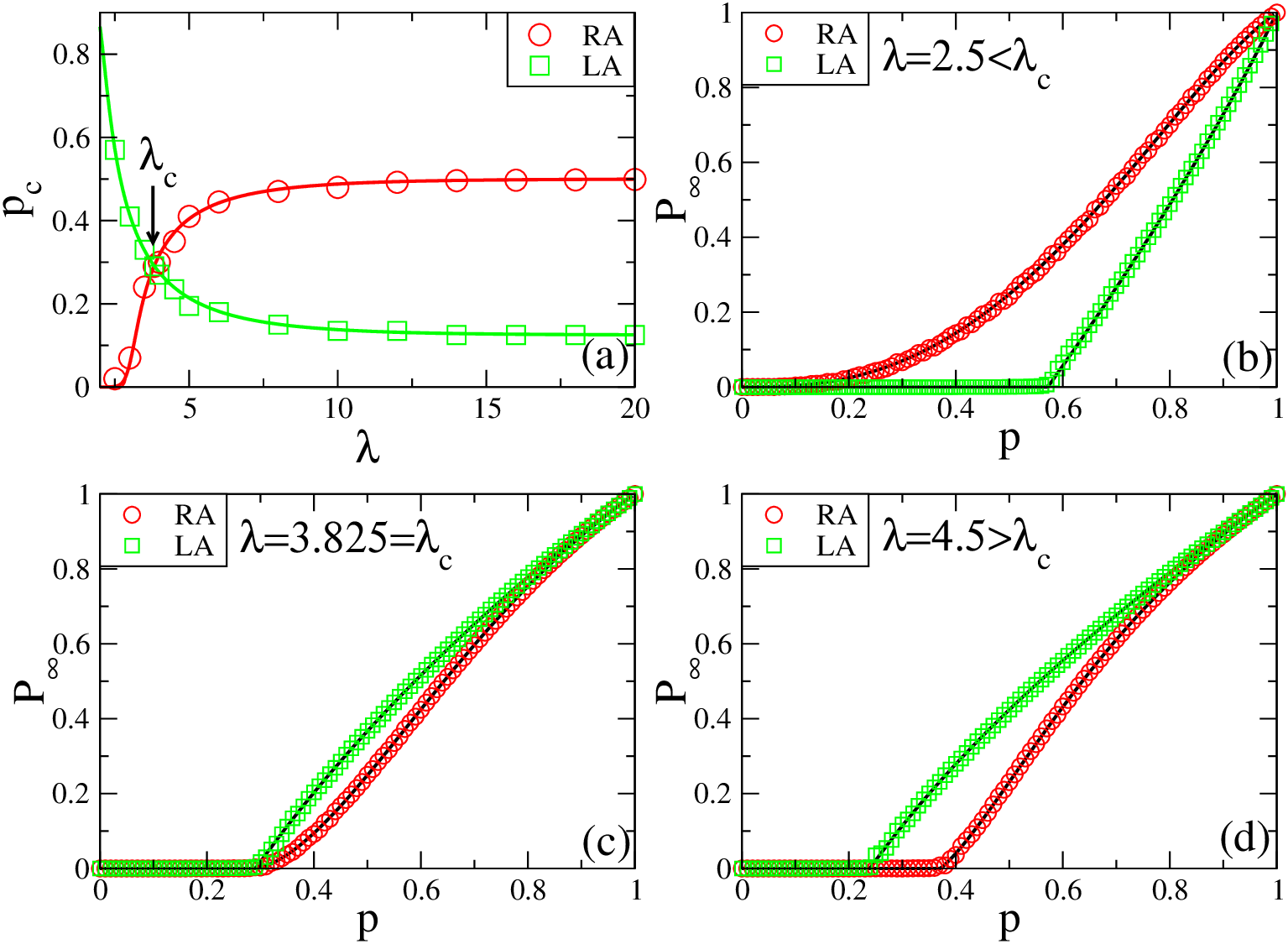}
\caption{Percolation properties for a SF network under localized attack (LA) and random attack (RA). Solid lines are from theory (Eq.~(\ref{gpx})) and symbols represent simulation results with $N=10^6$, $m=2$, and $\langle k\rangle=3$. (a) Critical threshold $p_c$ as a function of degree exponent $\lambda$. When $\lambda\rightarrow\infty$, the SF network converges to an RR network with $k_0=\langle k\rangle=3$, so $p_c (RA)\rightarrow 1/(k_0-1)=0.5$ and $p_c (LA)\rightarrow (k_0-1)^{-\frac{k_0}{k_0-2}}=0.125$, as confirmed in simulations. Note that for $2<\lambda \le3$, $p_c\rightarrow 0$ in the thermodynamic limit ($N\rightarrow \infty$) for random attack~\cite{Cohen2000}. (b) When $\lambda<\lambda_c$, the SF network is more vulnerable to
  localized attack compared to random attack. (c) When $\lambda=\lambda_c$, $p_c$ for localized attack and for random attack are equal. (d) When $\lambda>\lambda_c$, the SF network is more robust against localized attack compared to random attack.}
\label{fig_sf}
\end{center}
\end{figure}

For a SF network the degree distribution is $P(k)\sim k^{-\lambda}$
($m\le k\le M$), where $m$ and $M$ are the lower and upper bound of the
degree, respectively, and $\lambda$ is the power exponent. The critical
threshold $p_c$ and the size of the giant component $P_{\infty}(p)$ are
solved numerically by using the theoretical framework developed in
Eq.~(\ref{gpx}) (see Fig.~\ref{fig_sf}). We find that the degree
heterogeneity plays an important role in the robustness of SF networks
against localized attack. The critical threshold $p_c$ and the size of
the giant component $P_{\infty}(p)$ for the percolation transition of
the SF network under localized attack depends on $\lambda$. We find that
in a SF network there is a critical value $\lambda_c$ below which a
localized attack is significantly more severe than a random attack, but
when $\lambda>\lambda_c$ a random attack is more severe. Indeed, as seen
in Fig.~\ref{fig_sf}(a), for $\lambda<\lambda_c$, $p_c$ for a localized
attack is significantly higher than for a random attack. As $\lambda$
increases and the network becomes less heterogeneous, $p_c$ decreases
and the network becomes more robust against localized attacks. The
specific value of $\lambda_c$ depends on other parameters, such as $m$,
$M$, and $\langle k \rangle$. In Fig.~\ref{fig_sf}(b)$-$(d), we plot the
size of the giant component $P_{\infty}(p)$ as a function of $p$ and
compare the results of a localized attack with those of a random
attack. One intuitive explanation for the dependence of network
robustness on $\lambda$ is that, on the one hand, there is a higher
probability that higher degree nodes will be within the attacked hole,
which accelerates the fragmentation of the SF network; on the other,
only nodes on the surface of the attacked hole are connected to the
remaining network and contribute to its breakdown, which mitigates the
fragmentation process. The total impact of the localized attack is the
result of the competition between these two effects. As $\lambda$
increases and the SF network becomes less heterogeneous, the first
effect becomes less dominant and the network becomes more robust. Our
analytical analysis shows that for an ER network these two effects
always compensate each other and yield equal effects from both localized
attack and random attack. For an RR network, on the other hand, the
degrees are all the same and therefore only the second effect exists,
and the underlying network becomes more robust against localized attack
than against random attack.

We also investigate the robustness of real-world networks against localized attack and random attack using a peer-to-peer computer network~\cite{p2p} and a global airline route network~\cite{airline}. The real-world data proves the feasibility of our model, as shown in supplementary information.

To conclude, we have developed a mathematical framework for studying the
percolation of localized attacks on complex networks with an arbitrary
degree distribution. Using generating function methods, we have solved
exactly for the percolation properties of random networks under
localized node removal. Our results show that the effects of localized
attack and random attack on an Erd\H{o}s-R\'{e}nyi network are
identical. While a random-regular network is more robust against
localized attack than against random attack, the robustness of a
scale-free network depends on the heterogeneity of the degree
distribution. When $\lambda<\lambda_c$, the SF network is found to be
significantly more vulnerable with respect to localized attack compared
to random attack. When $\lambda>\lambda_c$, the opposite is true. Our
results can provide insight into understanding the robustness of complex
systems and facilitate the design of resilient infrastructures.

\section*{Acknowledgement}
We wish to thank ONR (Grant N00014-09-1-0380, Grant N00014-12-1-0548, Grant N62909-14-1-N019), DTRA (Grant HDTRA-1-10-1-0014, Grant HDTRA-1-09-1-0035), NSF (Grant CMMI 1125290), the European MULTIPLEX, CONGAS and LINC
projects, DFG, the Next Generation Infrastructure (Bsik) and the Israel
Science Foundation for financial support. We also thank the FOC program
of the  European Union for support.

\section*{References}
\begin{thebibliography}{99}

\bibitem{Albert2000} Albert R, Jeong H, and Barab$\acute{a}$si A L 2000 
{\it Nature (London)} {\bf 406} 6794; {\bf 406} 378

\bibitem{Cohen2000}Cohen R, Erez K, ben-Avraham D, and Havlin S 2000
{\it Phys. Rev. Lett.} {\bf 85} 4626

\bibitem{Callaway2000}Callaway D S, Newman M E J, Strogatz S H , and Watts D J 2000
{\it Phys. Rev. Lett.} {\bf 85} 5468 

\bibitem{Cohen2001} Cohen R, Erez K, ben-Avraham D, and Havlin S 2001
{\it Phys. Rev. Lett.} {\bf 86} 3682

\bibitem{Barabasi2002} Barab$\acute{a}$si  A L and Albert R 2002
{\it Rev. Mod. Phys.} {\bf 74} 47

\bibitem{Imre2005} Der$\acute{e}$yi \etal I 2005 
{\it Phys. Rev. Lett.} {\bf 94} 160202

\bibitem{Gallos2005} Gallos \etal L 2005 
{\it Phys. Rev. Lett.} {\bf 94} 188701

\bibitem{Newmanbook}  Newman M E J 2010 {\it Networks: An Introduction} (Oxford University Press, Oxford)

\bibitem{Bashan2011} Bashan A, Parshani R, and Havlin S 2011
{\it Phys. Rev. E } {\bf 83} 051127 

\bibitem{Sergey2010} Buldyrev S V \etal 2010
{\it Nature (London)} {\bf 464} 1025 


\bibitem{Parshani2010} Parshani R, Buldyrev S V, and Havlin S 2010
{\it Phys. Rev. Lett.} {\bf 105} 048701


\bibitem{Huang2011} Huang X, Gao J, Buldyrev S V, Havlin S, and Stanley H E 2011
{\it Phys. Rev. E } {\bf 83} 065101

\bibitem{Bashan2012} Bashan A, Bartsch R P, Kantelhardt J W, Havlin S, Ivanov P C 2012
{\it Nature Communications} {\bf 3} 702 

\bibitem{Jianxi2011} Gao J, Buldyrev S V,  Havlin S, and Stanley H E 2011
{\it Phys. Rev. Lett.} {\bf 107} 195701

\bibitem{Jianxi2012} Gao J, Buldyrev S V, Stanley H E, and Havlin S 2012
{\it Nature Physics} {\bf 8} 40 

\bibitem{Jianxi2013} Gao J , Buldyrev S V, Stanley H E, Xu X, and Havlin S 2013 
 {\it Phys. Rev. E} {\bf 88} 062816 

\bibitem{Brummitt2012} Brummitt C D, D'Souza R M, Leicht E A 2012
{\it Proc. Natl. Acad. Sci.} {\bf 109} 680 

\bibitem{Baxter2012} Baxter G J, Dorogovtsev S N, Goltsev A V, and Mendes J F F 2012
{\it Phys. Rev. Lett. } {\bf 109} 248701

\bibitem{Peixoto2012} Peixoto T P and Bornholdt S 2012
{\it Phys. Rev. Lett.} {\bf 109} 118703 

\bibitem{Cohenbook} Cohen R and Havlin S 2010
{\it Complex Networks, Structure, Robustness and Function} (Cambridge University Press, Cambridge)

\bibitem{Bundebook} Bunde A and Havlin S 1991 {\it Fractals and Disordered Systems} (Springer)

\bibitem{Staufferbook} Stauffer D and Aharony A 1994
{\it Introduction to Percolation Theory} (CRC Press)

\bibitem{Coniglio} Coniglio A 1982
{\it J. Phys. A: Math. Gen.} {\bf 15} 3829

\bibitem{Neumayer2009} Neumayer S, Zussman G, Cohen R,  Modiano E 2009 
{\it INFOCOM} IEEE 1566-1574

\bibitem{Berezin2013} Berezin Y, Bashan A, Danziger M M, Li D, and Havlin S 
{\it arXiv:} 1310.0996

\bibitem{BollobasBook} Bollob$\acute{a}$s B 1985
{\it Random Graphs} (London: Academic Press)

\bibitem{Newman2001} Newman M E J, Strogatz S H , and Watts D J 2001
{\it Phys. Rev. E} {\bf 64} 026118

\bibitem{Molly1995} Molly M and Reed B 1995
Random Struct. Algorithms {\bf 6} 161

\bibitem{Kalisky2006} Kalisky T, Cohen R, Mokryn O, Dolev D, Shavitt Y, and Havlin S 2006
{\it Phys. Rev. E} {\bf 74} 066108

\bibitem{JiaShao2009} Shao J, Buldyrev S V, Braunstein L A, Havlin S, and Stanley H E 2009
{\it Phys. Rev. E} {\bf 80} 036105 

\bibitem{Newman2002} Newman M E J 2002
{\it Phys. Rev. E} {\bf 66} 016128

\bibitem{p2p} Stanford Large Network Collection, Internet peer-to-peer network data. Available at http://snap.stanford.edu/data/.

\bibitem{airline} Openflight.org, Airport network data. Available at 
http://openflight.org/data.html.

\end{thebibliography}
\end{document}


\title{Percolation of localized attack on complex networks
  (Supplementary Information)} 

\author{Shuai Shao$^{1}$, Xuqing Huang$^1$, H. Eugene Stanley$^1$,
  and Shlomo Havlin$^{1,2}$} 

\affiliation{$^1$Center for Polymer Studies and Department of Physics,
Boston University, Boston, MA 02215, USA\\
$^2$Department of Physics,
Bar-Ilan University, Ramat-Gan 52900, Israel
}

\date{\today}

\maketitle

\section{Theoretical derivation of the generating function of the
  remaining network after localized attack} 

Consider a random network with arbitrary degree distribution $P(k)$,
which represents the probability of a node in the network to have $k$
links. The corresponding generating function is defined as

\begin{equation}
G_0(x)=\sum_{k=0}^{\infty}{P(k)x^k}.
\end{equation}
We separate the process of a localized attack into two stages: (i) at
the first stage, we remove all the nodes belonging to the attacked area
but keep the links connecting the removed nodes to the remaining nodes;
(ii) at the second stage, we remove those links. Now consider the degree
distribution $P_{p}(k)$ of the remaining nodes after the first
stage. Following Ref.~\cite{JiaShao2009} and letting $A_{p}(k)$ be the
number of nodes with degree $k$ in the remaining network, we have
\begin{equation}
P_p(k)=\frac{A_p(k)}{pN}. \label{Ppk}
\end{equation}
With one more node being removed, $A_p(k)$ changes as 
\begin{equation}
A_{(p-1/N)}(k)=A_p(k)-\frac{P_p(k)k}{\langle k(p)\rangle}, \label{Apk}
\end{equation}
where $\langle k(p)\rangle\equiv\sum{P_p(k)k}$. In the limit
$N\rightarrow\infty$, Eq.~(\ref{Apk}) can be presented in terms of a
derivative of $A_p(k)$ with respect to $p$,
\begin{equation}
\frac{dA_p(k)}{dp}=N\frac{P_p(k)k}{\langle k(p)\rangle}. \label{Ninfity}
\end{equation}
By differentiating Eq.~(\ref{Ppk}) with respect to $p$ and plugging it
into Eq.~(\ref{Ninfity}), we have
\begin{equation}
p\frac{dP_p(k)}{dp}+P_p(k)-\frac{P_p(k)k}{\langle
  k(p)\rangle}=0. \label{difPpk} 
\end{equation}
The solution of Eq.~(\ref{difPpk}) can be expressed as 
\begin{equation}
P_p(k)=P(k)\frac{f^k}{G_0(f)},
\end{equation} 
and the average degree of the remaining network is
\begin{equation}
\langle k(f)\rangle=\frac{fG^{'}_0(f)}{G_0(f)},
\end{equation}
where $f\equiv G^{-1}_0(p)$. Thus the generating function of $P_p(k)$ is
\begin{equation}
G_a(x)\equiv\sum_{k}{P_p(k)x^k}=\frac{G_0(fx)}{G_0(f)}.
\end{equation}
Now consider the second stage of removing the links of the remaining
nodes which lead to the removed nodes. The number of links belonging to
the nodes on the outer shell of the attacked hole, $L(f)$, can be
expressed as \cite{JiaShao2009},
\begin{equation}
L(f)=N(G^{'}_0(1)f^2-G^{'}_0(f)f). \label{lt}
\end{equation}
Since loops are allowed in the random network model, those links can be
connected either to the remaining nodes or to other nodes on the same
outer shell of the attacked hole. The number of links of the remaining
nodes which lead to the removed nodes is
\begin{equation}
\tilde{L}(f)=L(f)\frac{Np\langle k(f)\rangle}{Np\langle
  k(f)\rangle+L(f)}=N[fG^{'}_0(f)-\frac{G^{'}_0(f)^2}{G^{'}_0(1)}]. 
\end{equation}
The probability that a link in the remaining network will end at an
unremoved node is equal to  
\begin{equation}
\tilde{p}=1-\frac{\tilde{L}(f)}{pN\langle k(f)\rangle}=\frac{G^{'}_0(f)}{G^{'}_0(1)f}.
\end{equation}
Because the network is randomly connected, removing the links that end
at the removed nodes is equivalent to randomly removing a $1-\tilde{p}$
fraction of links of the remaining network. The generating function of
the remaining network after the random removal of a $1-\tilde{p}$
fraction of links is equal to \cite{Newman2002}
\begin{equation}
G^{p}_{0}(x)\equiv
G_a(1-\tilde{p}+\tilde{p}x)=
\frac{1}{G_0(f)}G_0[f+\frac{G^{'}_0(f)}{G^{'}_0(1)}(x-1)], \label{gpx}  
\end{equation}
where $f\equiv G^{-1}_0(p)$. Note that Eq.~(\ref{gpx}) is the generating
function of the degree distribution of the remaining network after a
localized attack.

\bigskip

\section{Robustness of real-world networks against localized attack}

\begin{figure}
\includegraphics[width=18cm,angle=0]{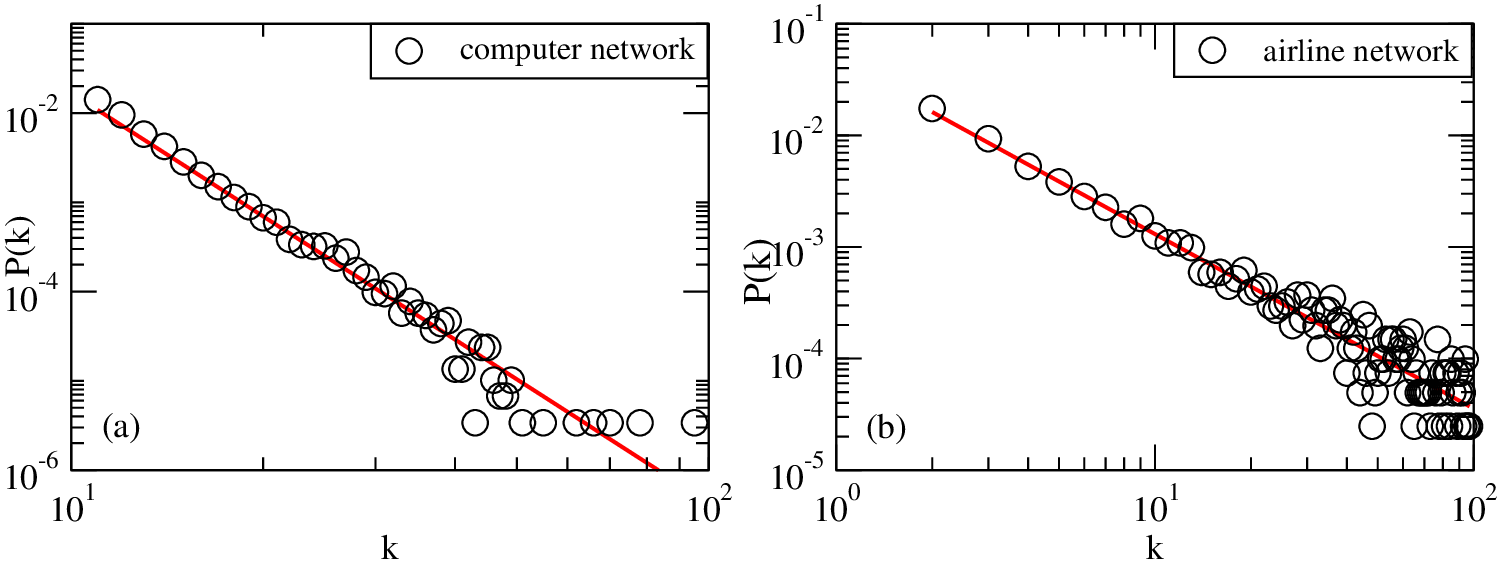}
\caption{Degree distribution of (a) the peer-to-peer computer network
  with $N$=62586, $\langle k\rangle=4.73$ and $\lambda=4.59$, and (b)
  the global airline route network with $N$=3308, $\langle
  k\rangle=12.2$ and $\lambda=1.57$. The degree distribution of both
  network approximately follow power law distribution.}
\label{fig_deg}
\end{figure}

\begin{figure}
\includegraphics[width=10cm,angle=0]{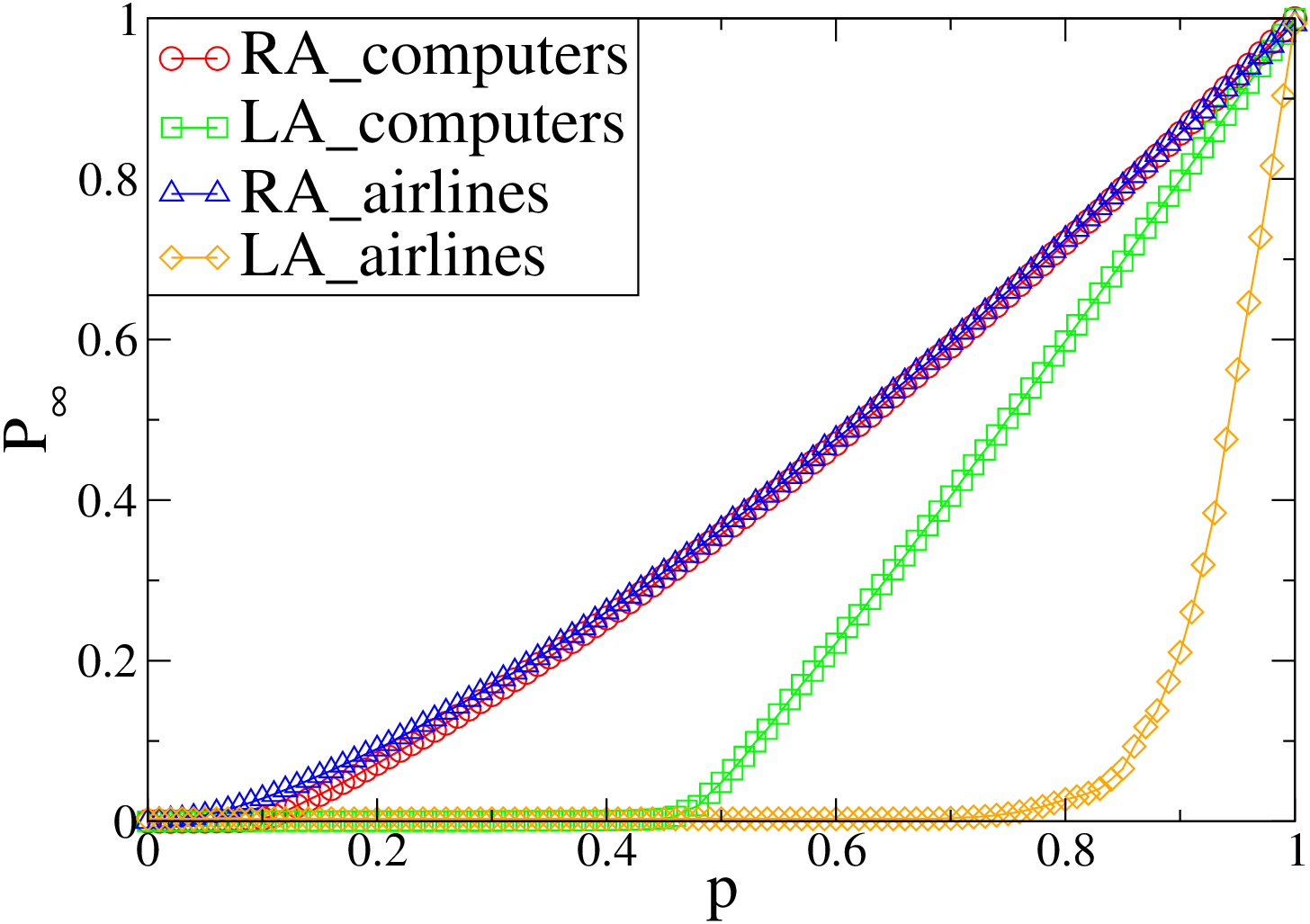}
\caption{ Robustness of real-world networks against localized attack (LA)
    and random attack (RA). A comparison of localized attack and random
  attack on a peer-to-peer computer network and a global airline route
  network~\cite{p2p, airline}. The size of the giant component $P_{\infty}(p)$ after locally
	attacking $1-p$ fraction of the whole network, versus
  $p$. The circles (red) and squares (green) represent simulation
  results of the peer-to-peer computer network ($N=62586$, $\langle
  k\rangle=4.73$ and $\lambda=4.59$) under random attack and localized
  attack respectively. The triangles (blue) and the diamonds (orange)
  represent simulation results of the global airline route network
  ($N=3308$, $\langle k\rangle=12.2$ and $\lambda=1.57$) under random
  attack and localized attack respectively. The simulation results are
  the average over 100 and 1000 realizations for the computer network
  and the airline network respectively. } 
\label{fig_realdata}
\end{figure}

We test and compare the robustness of real-world networks against
localized attack and random attack using a peer-to-peer computer
network~\cite{p2p} and a global airline route
network~\cite{airline}. The degree distributions of both networks 
approximately follow power law (see Figure~\ref{fig_deg}). Figure~\ref{fig_realdata} shows that, for both real-world networks, localized attack can collapse the network much more easily: a
node failure of 30\% in the global airline route network and 55\% in the
peer-to-peer computer network can disable the total network. When the
attack is random, however, a node failure of 98\% in the global airline
route network and 90\% in the peer-to-peer computer network must occur
before the network collapses. This shows that a localized attack is
significantly more harmful to real-world SF networks than a random
attack, supporting our theoretical results for SF networks with $\lambda<\lambda_c$.

\vfill\eject

